\documentclass[prl,twocolumn,superscriptaddress,floatfix]{revtex4}

\usepackage{amsmath}
\usepackage{amstext}
\usepackage{amssymb}
\usepackage{latexsym}
\usepackage[dvips]{graphicx}

\newcommand{\beq}{\begin{equation}}
\newcommand{\eeq}{\end{equation}}
\newcommand{\barr}{\begin{eqnarray}}
\newcommand{\earr}{\end{eqnarray}}
\newcommand{\ket}[1]{\left\vert#1\right\rangle}
\newcommand{\bra}[1]{\left\langle#1\right\vert}
\newcommand{\Ham}{\mathcal H}

\begin{document}
\title{Mott insulating and glassy phases of polaritons in 1D arrays of coupled cavities}

\author{Davide Rossini}
\affiliation{NEST-CNR-INFM \& Scuola Normale Superiore,
  Piazza dei Cavalieri 7, I-56126 Pisa, Italy}
\author{Rosario Fazio}
\affiliation{International School for Advanced Studies (SISSA),
  Via Beirut 2-4, I-34014 Trieste, Italy}
\affiliation{NEST-CNR-INFM \& Scuola Normale Superiore,
  Piazza dei Cavalieri 7, I-56126 Pisa, Italy}

\begin{abstract}
By means of analytical and numerical methods we analyze the phase diagram of
polaritons in one-dimensional coupled cavities. We locate the phase boundary, 
discuss the behavior of the polariton compressibility and visibility fringes 
across the critical point, and find a non-trivial scaling of the phase boundary 
as a function of the number of atoms inside each cavity. We also predict the 
emergence of a {\em polaritonic glassy phase} when the number of atoms fluctuates 
from cavity to cavity.
\end{abstract}

\maketitle

Over the past two decades a considerable understanding of the physics of 
strongly interacting systems has been gained by a judicious design of 
controlled many-body systems. Successful examples of this sort were optical 
lattices or Josephson junction arrays (see the reviews~\cite{lewenstein06,fazio01}). 
The recent proposals~\cite{plenio06,hollenberg06,angelakis06} to realize a 
Mott phase of polaritons have paved the way to use coupled cavities for the 
study of strongly correlated phenomena in a controlled way. The rich scenario 
which emerges in these systems stems from the interplay of two effects. Light-matter 
interaction inside the cavity leads to a strong effective Kerr nonlinearity between 
photons. By controlling the atomic level spacings and the photonic resonance 
frequency inside the cavity, it is possible to achieve a photon blockade 
regime~\cite{imamoglu97,rebic99,kim99,birnbaum05}, thus suppressing photon 
fluctuations in each cavity. On the other hand, photon hopping between 
neighboring cavities favours delocalization thus competing with photon 
blockade. 
Coupled cavities can be realized in a wide range of physical systems, from nanocavities in 
photonic crystals~\cite{akahane03} to Cooper pair boxes in superconducting 
resonators~\cite{wallraff04}. It is therefore possible to study a whole 
new class of strongly interacting systems that, for the first time, can be addressed 
and measured locally.

The polariton Mott insulator has been predicted in two cases. Hartmann {\em 
et al.}~\cite{plenio06} discussed a cavity doped with $N$ four-level systems in the 
limit of large $N$, while Angelakis {\em et al.}~\cite{angelakis06} and Greentree 
{\em et al.}~\cite{hollenberg06} analyzed the Jaynes-Cummings~\cite{jcreview93} model 
as a scheme for the light-matter interaction; in this last case, an experimental
proposal has been also devised by Neil Na {\em et al.}~\cite{yamamoto07}.
Hartmann {\em et al.} found a mapping onto a Bose-Hubbard model~\cite{fisher89} for 
the polaritons in the limit of large number of atoms and large detuning. In the 
other case, the phase boundary was evaluated at a mean field 
level for one~\cite{hollenberg06} and many~\cite{yamamoto07} atoms in cavity.
The exact phase diagram has not been worked out so far; this is what 
we accomplish in this work for the one-dimensional case. By means of numerical simulations 
and analytical calculations we are able to locate the phase boundary and its non-trivial
scaling as a function of the number of atoms in the cavity. 
Furthermore we consider the case where the number of atoms fluctuates in each cavity
and we show that this leads to the existence of a polariton glass. 

The Hamiltonian for the system composed by an array of $L$ identical
coupled cavities is given by the local Hamiltonian
on each cavity and the photon hopping term between different cavities:
\beq
\Ham = \sum_{i=1}^L \Ham_i^{(a)} - t \sum_{\left< i,j \right>}
\left( a^\dagger_i a_j + a^\dagger_j a_i \right)
- \mu \sum_{i=1}^L n_i \, .
\label{eq:fullham}
\eeq
As in~\cite{hollenberg06,yamamoto07}, we add a chemical potential $\mu$. 
In the previous expression $t$ is the nearest-neighbor inter-cavity 
photon hopping, and $a_i$ is the photon annihilation operator in the $i$-th cavity;
the local contribution $\Ham_i^{(a)}$ describes the light-matter interaction.
We will consider the following two models.

\paragraph{Model I}
A collection of $N$ two-level systems which interact with photons 
via a Jaynes-Cummings coupling
%\beq
$
\Ham_i^{(I,a)} = \epsilon ( S^z_i + \frac{N}{2} )
+ \omega a^\dagger_i a_i + \beta ( S^+_i a_i + S^-_i a^\dagger_i )
$,
%\eeq
where we have defined the spin operators
$S^{\alpha}_i = \sum_{j=1}^N \sigma_{j,i}^{\alpha}\,$ ($\alpha = \pm, z$) and
$ \sigma_{j,i}^{\pm}$ are the atomic raising/lowering operators for the $j$-th atom,
$\epsilon$ denotes the transition energy between the two atomic
levels, $\omega$ is the resonance frequency of the cavity, and $\beta$
is the atom-field coupling constant ($\epsilon, \, \omega, \, \beta > 0$).
The total number of atomic plus photonic excitations and the total
atomic spin $S^2_i$ on each site are conserved quantities.
The ground state is always in the subspace of maximum spin, $S=N/2$.

\paragraph{Model II}
In the Jaynes-Cummings model at a large detuning $\Delta$, when the atomic spontaneous 
emission is minimized, also the strength of nonlinearities is weakened.
In order to overcome this problem, a different scheme involving four-level atoms, 
has been proposed~\cite{schmidt96} producing a large Kerr nonlinearity with virtually 
no noise. In the interaction picture, in electric dipole and rotating wave 
approximations the model reads
%\begin{eqnarray}
$
  \Ham^{(II,a)}_i = \delta S^{33}_i + \Delta S^{44}_i
  + \Omega ( S^{23}_i + S^{32}_i )
  + g_{1} ( S^{13}_i  \, a^\dagger  + S^{31}_i \, a )
  +g_{2} ( S^{24}_i  \, a^\dagger  + S^{42}_i \, a ) \nonumber
$,
%\end{eqnarray}
having defined the global atomic raising and lowering operators
$S^{lm} = \sum_{j=1}^N \ket{l}_j \bra{m}$;
$\sigma^{lm}_{j} = \ket{l}_j \bra{m}$ are the atomic raising and lowering
operators ($l \neq m$), or energy level populations ($l = m$)
for the $j$-th atom.
The transition $\ket{3}_j \to \ket{2}_j$ is driven by a classical coupling
field with Rabi frequency $\Omega$; the cavity mode of frequency
$\omega_{cav}$ couples the $\ket{1}_j \to \ket{3}_j$ and
$\ket{2}_j \to \ket{4}_j$ transitions with coupling constants $g_1$ and $g_2$;
the parameters $\delta$ and $\Delta$ account for the detunings of levels
$3$ and $4$ respectively.
The atomic part of the system wavefunction for the $i$-th cavity can be
fully characterized by the number of atoms in each of the four possible states:
$\{ \ket{n_1, n_2, n_3, n_4} \}$, with $\sum_{i=1}^4n_i = N$.
The total number of photons
plus the number of atomic excitations in the whole system
(where states $\ket{2}_j, \ket{3}_j$ count
for one excitation, while $\ket{4}_j$ counts for two excitations),
is a conserved quantity.
Hereafter we assume $g_1 \simeq g_2 \equiv g$ and define
the relative atomic detuning $\delta_w \equiv \Delta - \delta$.

\paragraph{Mott insulator}
The phase diagram of the coupled cavity system is characterized by two 
distinct phases~\cite{plenio06,hollenberg06,angelakis06}:
the Mott Insulator (MI) is surrounded by the
Superfluid (SF) phase. In the MI polaritons are localized
on each site, with a uniform density $\rho \equiv n^{pol} /L$,
where $n^{pol}$ is the total number of polaritons in a system
of $L$ cavities; there is a gap in the spectrum,
and the compressibility $\kappa \equiv \partial \rho / \partial \mu$ vanishes.
A finite hopping renormalizes this gap, which eventually vanishes at $t^*$.
The phase boundaries between the two phases can thus be determined by
evaluating, as a function of the hopping, the critical values of $\mu$
at which the gap vanishes. Our data have been obtained by means of the Density 
Matrix Renormalization Group (DMRG) algorithm with open boundary conditions~\cite{footnote1}.
In numerical calculations, the Hilbert space for the on-site Hamiltonian is
fixed by a maximum number of admitted photons $n^{phot}_{\rm max}$.
We chose $n^{phot}_{\rm max} = 6$ for model $I$ and $n^{phot}_{\rm max} = 4$
for model $II$; we also retained up to $m=120$ states in the DMRG procedure,
such to guarantee accurate results, and checked that our data are not
affected by increasing $n^{phot}_{\rm max}$.
We simulated systems with up to $L=128$, and up to $N=5$
atoms per cavity~\cite{footnote2}; the asymptotic values in the thermodynamic limit
have been extracted by performing a linear fit in $1/L$.
By combining these results with strong coupling perturbation theory~\cite{freericks96} 
we were able to locate the phase boundaries for all values of $N$.
Most of this paper is devoted to the case $\epsilon=\omega$ for model $I$
and $\delta=\Delta=0$ for model $II$.
These regimes could not be accessed by the perturbative approach
of Ref.~\cite{plenio06}.

Let us start with zero photon hopping ($t=0$).
For model $I$, at fixed $N$, there exists a value $\delta^*_I$ of the detuning
$\delta_I \equiv \omega - \epsilon$ such that, for $\delta_I > \delta_I^*$, the width of
the lobe with a polaritonic density $\rho = N$ is greatly enhanced
with respect to the other lobes. We estimate $\delta^*_I$ numerically
and find a scaling $\delta^*_I \sim \sqrt{N}$.
For model $II$, at a given relative atomic detuning $\delta_\omega >0$ the situation
is similar to model $I$, where the resonating lobe with $\rho = N$ is much larger than 
the other lobes, if $\delta < \delta^*$. In the opposite case, $\delta_\omega <0$,
some of the lobes disappear.

%%%%%%%%%%%%%%%%%%%%%%%%%%%%%
\begin{figure}
  \begin{center}
    \includegraphics[scale=0.305]{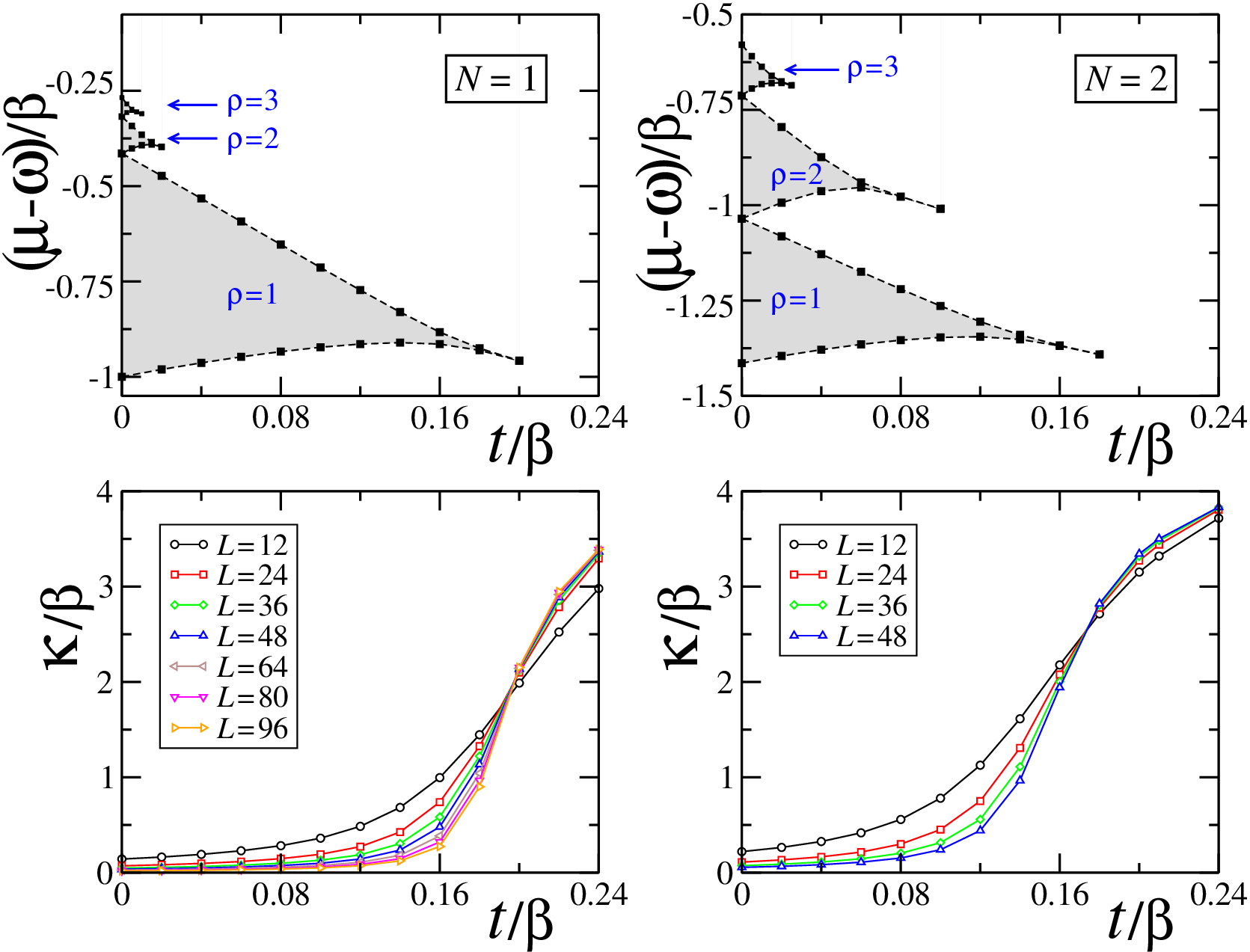}
    \caption{(Color online) Upper panels: Phase diagram for the
      Hamiltonian model $I$, with $N=1,2$ atoms inside each cavity
      at $\epsilon = \omega$.
      Lower panels: System compressibility $\kappa$ for the first
      lobe (i.e., $\rho = 1$), for different system sizes $L$,
      with $N=1$ (left) and $N=2$ (right).}
    \label{fig:PhaseDiag_I}
  \end{center}
\end{figure}
%%%%%%%%%%%%%%%%%%%%%%%%%%%%%

\begin{figure}
  \begin{center}
    \includegraphics[scale=0.305]{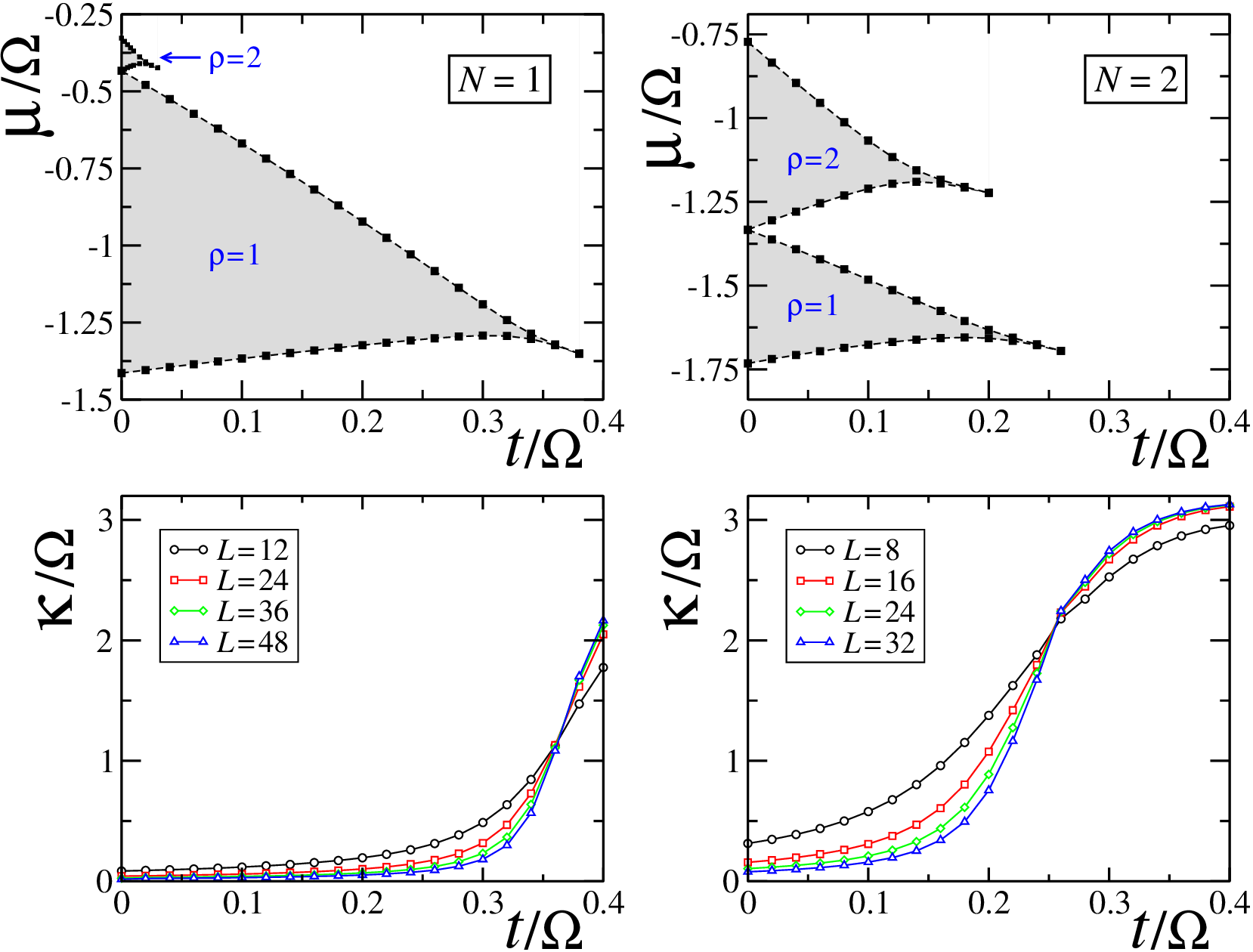}
    \caption{(Color online) Upper panels: Phase diagram of model $II$.
      The detuning parameters are set to zero and $g / \Omega = 1$.
      Lower panels: System compressibility $\kappa$  for the first
      lobe, for different system sizes $L$, with $N=1$ (left) and $N=2$ (right).}
    \label{fig:PhaseDiag_II}
  \end{center}
\end{figure}

For model $I$, numerical data at finite photon hopping for different
values of $N$ are shown in Fig.~\ref{fig:PhaseDiag_I};
the phase diagram of model $II$ is shown in Fig.~\ref{fig:PhaseDiag_II}.
Several interesting features emerge in the structure of the lobes. In both models,
for fixed $N$, contrary to Bose-Hubbard model, the critical values
$t^*$ of the hopping strength at which the various lobes shrink in a point
are not proportional to the lobe width at $t=0$. Furthermore, the ratio between the
upper and the lower slopes of the lobes at small hopping is greater than the one
predicted in Ref.~\cite{freericks96}; this discrepancy disappears on increasing
the number of atoms inside the cavity.
In terms of an effective Bose-Hubbard model, this may be understood as a
{\em correlated hopping} of the polaritons, i.e., the hopping depends
on the occupation of the cavity. 

%%%%%%%%%%%%%%%%%%%%%%%%%%%%%
%%%%%%%%%%%%%%%%%%%%%%%%%%%%%

\begin{figure}
  \begin{center}
    \includegraphics[scale=0.26]{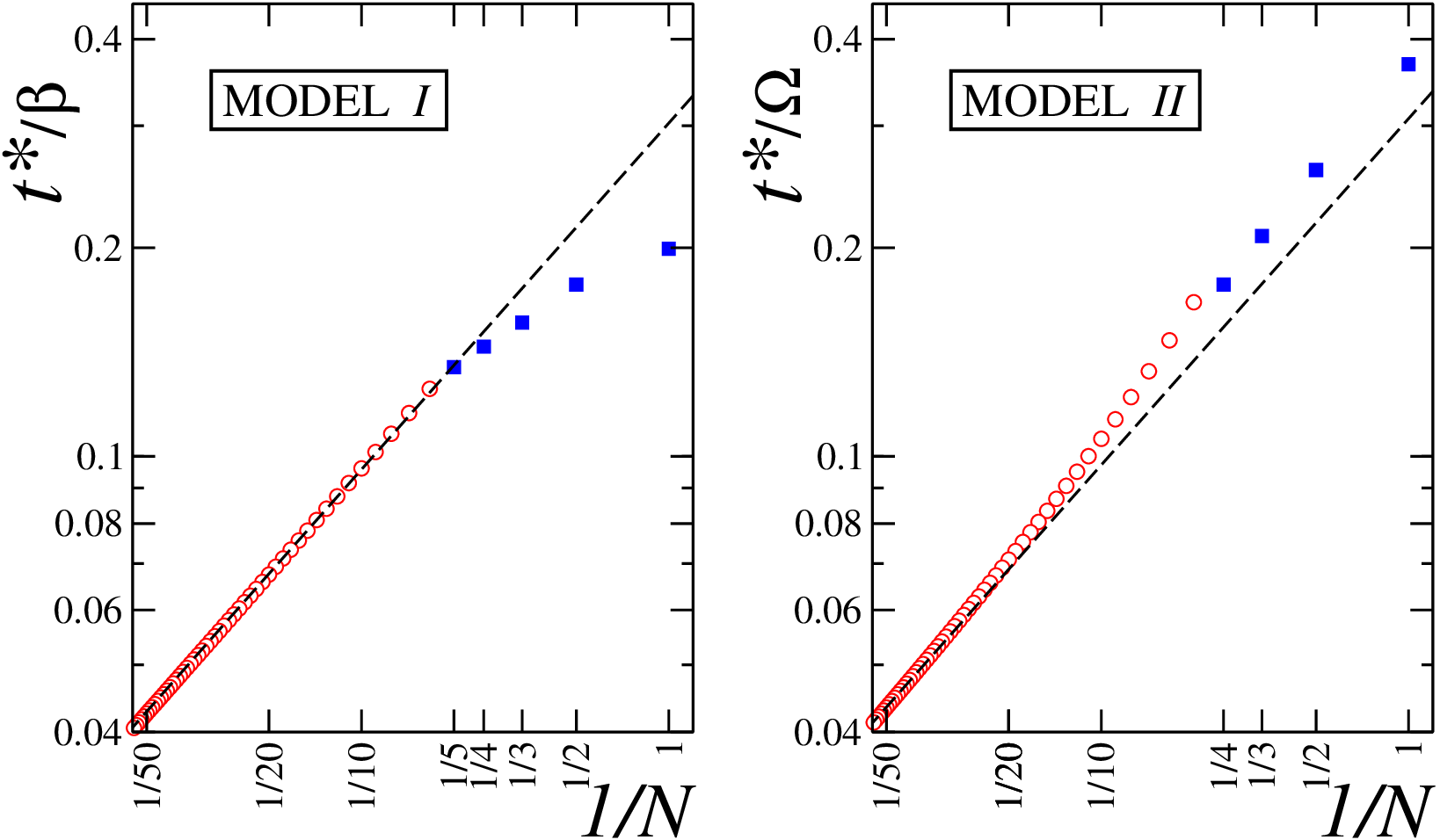}
    \includegraphics[scale=0.30]{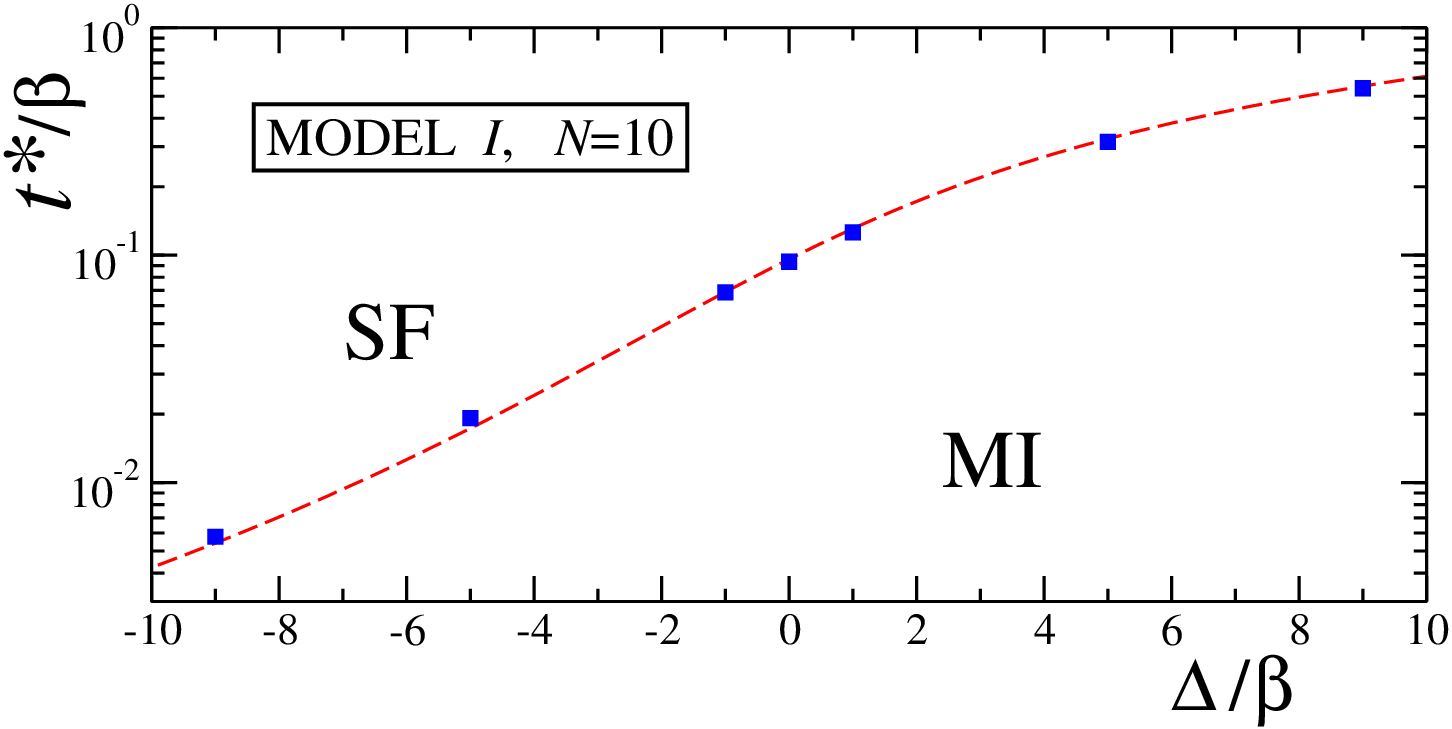}
    \caption{(Color online) Upper graphs: Critical hopping $t^*$ in the
      first lobe as a function of the number $N$ of atoms inside each cavity,
      for the two different models with all the detunings set to zero.
      Dashed black lines indicate a behavior
      $t^* \sim 1/\sqrt{N}$ and are plotted as guidelines.
      Lower graph: Critical hopping $t^*$ for model $I$ with $N=10$ atoms
      per cavity~\cite{footnote3}, as a function of the relative detuning $\Delta$.
      Blue squares have been evaluated with the DMRG.
      Red data are estimates obtained from the effective on-site polaritonic
      repulsion $U_{\rm eff}$ at zero hopping, with filling $\rho = 1$.}
    \label{fig:MI-SF}
  \end{center}
\end{figure}

A more detailed analysis of the transition at filling one can be performed 
by considering the compressibility.
In our DMRG simulations we fix the total number $n^{pol}$ of excitations in
the system, thus fixing the polariton density $\rho$ inside each cavity
in the insulating regime.
In the lower panels of Figs.~\ref{fig:PhaseDiag_I},~\ref{fig:PhaseDiag_II}
we plot $\kappa$ in the first insulating lobe
(with a polariton density $\rho=1$) as a function of $t$ for different sizes
of the system, for the models $I$ and $II$.
By exploring the mapping to the Bose-Hubbard model we construct
the full curve $t^*$ versus $N$.
The effective repulsive interaction $U_{\rm eff}$ between polaritons,  
at $\rho=1$, is given by
$U_{\rm eff} (1) \equiv \partial^2 E (n)/ \partial n^2 \vert_{n=1}$
(where $E(n)$ is the ground state energy of Hamiltonian $\Ham_i$
with $n$ polaritons), that is exactly the opening of the first lobe at $t=0$.
For model $I$ it is possible to give an exact analytic formula:
$U_{\rm eff} (1) = 2 \sqrt{N} [ 1 - \sqrt{1- 1/(2N)} ]$, while for model $II$ it 
can be evaluated numerically. As $N$ increases, $U_{\rm eff} (1)$ decays to zero;
for both models $U_{\rm eff} (1) \sim 1/\sqrt{N}$ as far as $N \to \infty$.
Moreover the effective repulsion depends on the number of polaritons,
contrary to the Bose-Hubbard model; this dependence weakens, and eventually
vanishes in the limit $n \ll N$. Therefore the mapping becomes accurate when 
$N$ increases. The polaritonic hopping $t_{\rm eff}$ can be obtained
by performing a strong coupling expansion in $t$.
For model $I$ we found that $t^* \sim 2 \, t^*_{\rm eff}$,
while for model $II$ we get $t^* \sim 2 \, \frac{N+1}{N} \, t^*_{\rm eff}$.
The critical hopping is then obtained using the value for the critical point
$t^*_{\rm eff} / U_{\rm eff} \simeq 0.3$~\cite{kuhner00}.
Figure~\ref{fig:MI-SF} (upper part) displays both numerical (blue squares)
and analytical estimates (red circles) for the two models. 
This analysis shows that the Bose-Hubbard model provides a good description
already for $N \sim 10$.
A study of the dynamics is needed to further strengthen this observation.
We point out that, in experimental realizations, the parameter that can be
changed to cross the transition is the detuning.
For model $I$ this is shown in the lower part of Fig.~\ref{fig:MI-SF}.

%%%%%%%%%%%%%%%%%%%%%%%%%%%%%
%%%%%%%%%%%%%%%%%%%%%%%%%%%%%

\paragraph{Visibility of photon interference}
The phase transition can be detected by analyzing the phase coherence 
of photons~\cite{sun07}, in a way similar to what has been done for 
the Bose-Hubbard model~\cite{bloch05,sun07}.
The interference pattern of the photonic density is proportional to the
photon number distribution ${\cal S}$ in the momentum space:
$
{\cal S} (k) = \frac{1}{L} \sum_{j,l=1}^L e^{2 \pi i k (j-l) / L}
\langle a_j^\dagger a_l \rangle \,.
$
The visibility of interference fringes can then be defined as
$
{\cal V} = ({\cal S}_{\rm max} - {\cal S}_{\rm min})/(
{\cal S}_{\rm max} + {\cal S}_{\rm min}) \, .
$
%%%%%%%%%%%%%%%%%%%%%%%%%%%%%
\begin{figure}
  \begin{center}
    \includegraphics[scale=0.3]{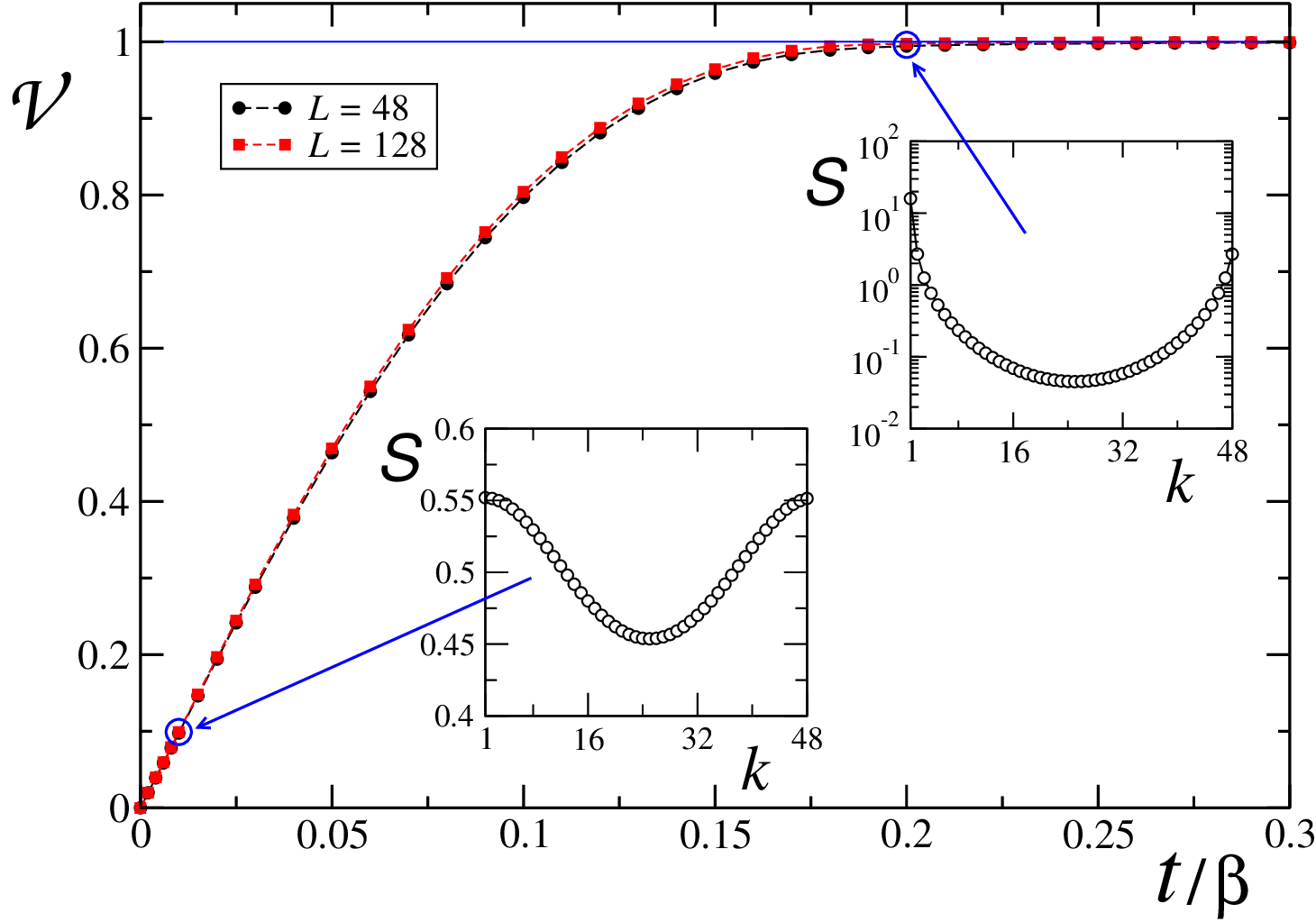}
    \caption{(Color online) Photonic visibility ${\cal V}$ in the first insulating lobe
      ($\rho = 1$) as a function of the hopping strength $t/\beta$, for model $I$.
      Here we considered the case $N=1$, $\Delta=0$.
      Insets: The photon number correlation function ${\cal S}(k)$
      for $L=48$ in the deep insulating regime ($t / \beta =0.01$),
      and at the edge of the superfluid regime ($t / \beta =0.20$).}
    \label{fig:Visib1_I}
  \end{center}
\end{figure}
%%%%%%%%%%%%%%%%%%%%%%%%%%%%%
%
The visibility is strictly zero only in the limit $t=0$, where
the interference pattern ${\cal S}$ is constant.
When $t$ is increased, the visibility itself increases, until it
saturates to the maximum value ${\cal V} = 1$ in the superfluid regime.
This description is well adapted even for photon coherence in our hybrid
light-matter system, as shown in Fig.~\ref{fig:Visib1_I}. 
The existence of different phases can also be detected by measuring
fluctuations in the number of polaritons, as discussed in~\cite{plenio06}.

\paragraph{Polaritonic glass phase}
Up to now we have assumed that the number of atoms in each cavity was
constant and equal to $N$.
In certain implementations this requirement might be 
demanding. Here however we consider this problem from a different 
perspective and show that, when $N$ changes from cavity to cavity, it leads
to the emergence of a {\em polariton glass}. Following~\cite{fisher89}, this
phase is characterized by a finite compressibility,
gapless excitation spectrum, and zero superfluid density. 

Random fluctuations in the number of atoms per cavity lead to 
disorder in the on-site light-matter interaction strength. This effect 
can cause significant consequences only in the limit of large $N$ (we quantify 
this statement below) where the mapping onto a Bose-Hubbard model applies.
A Bose glass phase has been originally predicted~\cite{fisher89} as a function 
of disorder in the chemical potential. Recently it has been shown 
that fluctuations in the on-site repulsion lead to a Bose glass as
well~\cite{gimperlein05}.

%%%%%%%%%%%%%%%%%%%%%%%%%%%%%
\begin{figure}
  \begin{center}
    \includegraphics[scale=0.33]{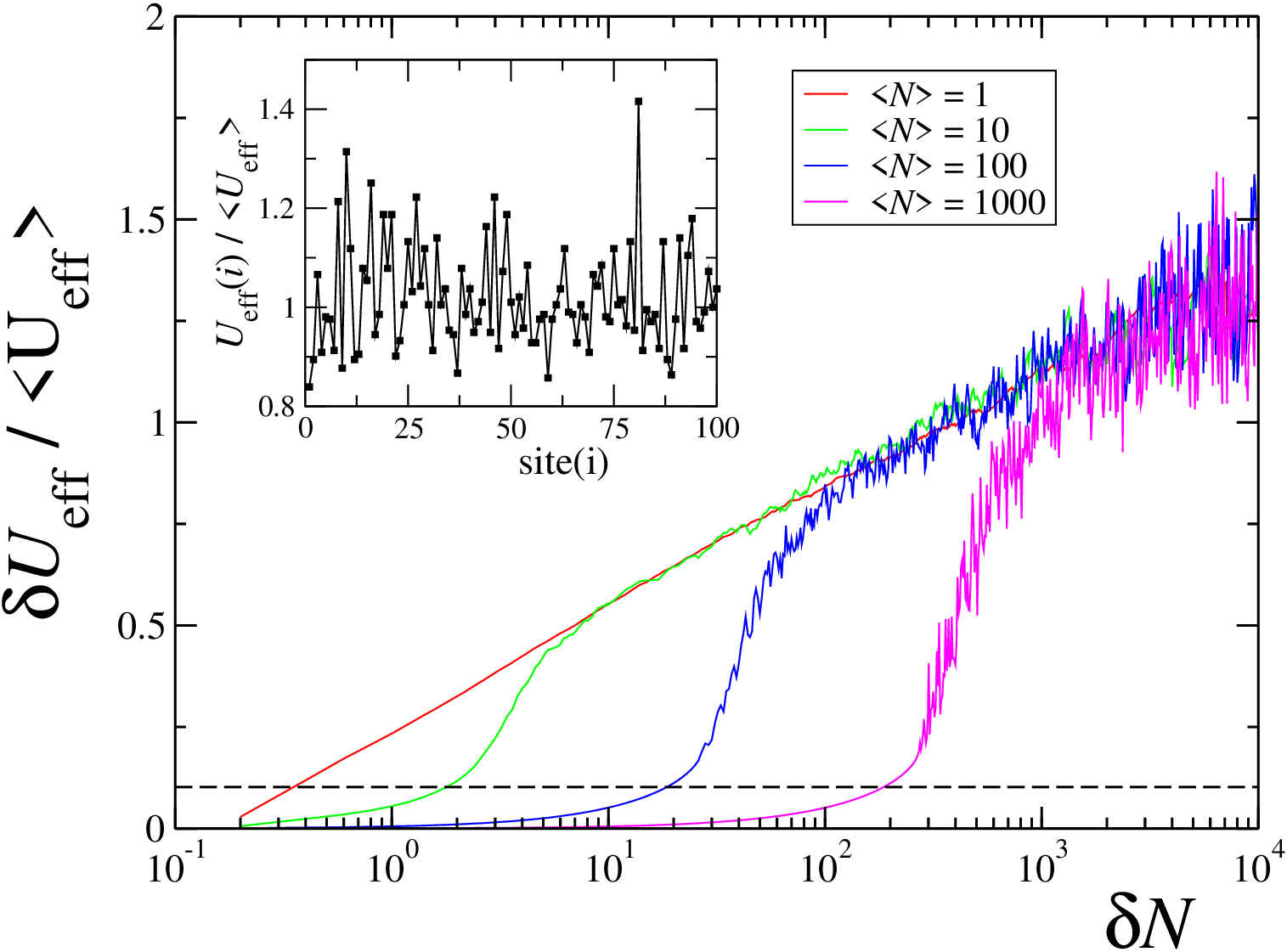}
    \caption{(Color online) Relative standard deviation of the effective
      on-site interaction strength $U_{\rm eff}$ averaged over $L=10^4$
      cavities, as a function of the atom number fluctuations $\delta N$; 
      from left to right $\left< N \right> = 1, \, 10, \, 100, \, 1000$.
      Dashed line indicates an effective variation of the interaction strength
      equal to the standard deviation of a random interaction uniformly
      distributed in the interval
      $U_{\rm eff} (i) / \left< U_{\rm eff} \right> \in [-\epsilon,+\epsilon]$
      with $\epsilon=0.25$.
      In the inset an example of the variation on the on-site effective
      interaction is shown; $\left< N \right> = 100$, $\delta N = 20$.}
    \label{fig:Glass_I}
  \end{center}
\end{figure}
%%%%%%%%%%%%%%%%%%%%%%%%%%%%%
We take advantage of the results obtained in~\cite{gimperlein05} and the mapping
to an effective Bose-Hubbard model  to give 
a detailed estimate for the width of the polariton glass phase. The key 
to our finding is to relate fluctuations in the number of atoms to 
fluctuations in the on-site repulsion $U_{\rm eff}$.
We suppose that each $N_i$ is a random discrete Gaussian variable
with a mean value $\left< N \right>$, and a standard deviation $\delta N$.
Figure~\ref{fig:Glass_I} displays fluctuations in the effective on-site
repulsion $\delta U_{\rm eff}$ as a function of the fluctuations
in the number of atoms, while in the inset of Fig.~\ref{fig:Glass_I} we show 
an example of such a variation, with $\left< N \right> = 100$, $\delta N = 20$,
in the case of model $I$.
We can then use numerical data of Ref.~\cite{gimperlein05}
to estimate where the polaritonic glass phase can be observed.
It has been shown that in the Bose Hubbard model, for a relative
interaction $\delta U_{\rm eff} / \left< U_{\rm eff} \right>$ uniformly
distributed in an interval of length $2\epsilon = 0.5$, a $L=200$ sites
system at filling $\rho =1.01$ exhibits a Bose glass phase for
$0.078 \lesssim t_{\rm eff} / \left< U_{\rm eff} \right> \lesssim 0.133$.
In our model with the same polaritonic filling,
if we take, e.g., $\left< N \right> =100$ particles per cavity
and we choose $\delta N \simeq 19$ (such to have $\epsilon = 0.25$),
a polaritonic glassy phase should be visible, in the case of model $I$, for
$
7.51 \times 10^{-3} \lesssim t/\beta \lesssim 1.401 \times 10^{-2} \;\;.
$
The situation is qualitatively the same for model $II$.
The polariton glass may be observed by measuring, at fractional fillings, 
the photon visibility as a function of the disorder.
Anyway, a complete characterization of the polariton glass requires
a study of its dynamical behavior.

\paragraph{Concluding remarks}
In this work we have discussed in details the equilibrium properties
of a chain of coupled cavities. The results we presented here are general,
in the sense that they apply to all the systems described
by Eq.~\eqref{eq:fullham}.
An important issue is to understand the effect of decoherence, decay,
cavity losses, \ldots , which occur in these systems.
These aspects have to be discussed for each implementation
(see for example~\cite{plenio06,wallraff04}), together with the effect
of an additional possible external driving.
The combined presence of dissipation and external driving may lead
to a change of the universality class of the transition, or 
to new phenomena associated to non-equilibrium phases.

We acknowledge useful discussions with A. Tredicucci and support 
by MIUR-PRIN and EC-Eurosqip.


\begin{thebibliography}{99}
\bibitem{footnote2}
  In model $II$, for $N=3,4$, only states with at most
  three atomic excitations and $n^{phot}_{\rm max} = 3$
  were considered.
\bibitem{footnote3}
  For $N=10$ we limited the total polariton number per site to $n=4$.
\end{thebibliography}
\end{document}